\documentstyle[prl,aps]{revtex} 
\begin{document}
\draft
\twocolumn[\hsize\textwidth\columnwidth\hsize\csname @twocolumnfalse\endcsname
\title{On weak vs.\ strong universality 
in the two-dimensional\\ 
random-bond Ising ferromagnet}
\author{F. D. A. Aar\~ao Reis, S. L. A. de Queiroz,
and Raimundo R. dos Santos}
\address{
Instituto de F\'\i sica, Universidade Federal Fluminense, Avenida Litor\^anea s/n, 24210-340 Niter\'oi RJ, Brazil}
\date{\today}
\maketitle
\begin{abstract}
We address the issue of universality in two-dimensional disordered Ising
systems, by considering long, finite-width strips of ferromagnetic Ising 
spins with randomly distributed couplings.
We calculate the free energy and spin-spin correlation functions
(from which averaged correlation lengths, $\xi^{ave}$, are computed)
by transfer-matrix methods.
An {\it ansatz} for the size-dependence of logarithmic corrections
to $\xi^{ave}$ is proposed.
Data for both random-bond and site-diluted systems show that pure system
behaviour (with $\nu=1$) is recovered if these corrections
are incorporated, discarding the weak--universality scenario.
\end{abstract}

\narrowtext
\vskip2pc]

It is well known that the Harris criterion~\cite{har}, for the relevance or irrelevance of
weak disorder upon critical behaviour at a phase transition, is
inconclusive for the two-dimensional Ising model
where the specific heat of the pure
system diverges logarithmically at the critical point. A great deal of
effort has been dedicated to elucidating the properties of disordered
versions of this model\cite{sst}. Both weak and strong disorder have been
considered. Early proposals implying strong deviations from pure system
behaviour, such as a peculiar exponential decay of critical correlations
with distance and a magnetization exponent $\beta = 0$~\cite{dotsenko} have
been ruled out by extensive numerical simulations\cite{sst}.
Though nowadays there seems to be general agreement that no such drastic
changes are expected to arise from disorder in this case, 
two main pictures have 
taken hold recently, which seem
to be mutually excludent. The first, typically represented by the work
of Heuer~\cite{heuer} and of Talapov and Shchur\cite{talapov} maintains
that the critical behaviour is unaffected by disorder (apart from possible
logarithmic corrections, which though not explicitly considered in Refs.
\onlinecite{heuer} and \onlinecite{talapov}, fit in with similar overall 
conclusions\cite{sst}); this view agrees with early numerical work
on magnetization moments\cite{dss}.  
According to the second view~\cite{kim,kuhn}, critical quantities such
as the zero-field susceptibility and correlation length display power-law
singularities, with the corresponding exponents $\gamma$ and $\nu$ changing continuously with disorder; 
however, this variation is such that the ratio $\gamma/\nu$ is kept constant 
at the pure system's  value (the so-called {\it weak universality} 
scenario\cite{suzuki}). 

The present work aims at shedding light into this controversy,
by means of strip calculations which, since
the work of Nightingale~\cite{Nightingale76,fs2} connecting 
Finite-Size Scaling~\cite{Fisher71,fs1} and Renormalization
Group ideas, have proved to be among the most accurate
techniques to extract critical points and exponents for non-random
systems in two dimensions. 
Extensions of this approach to random systems 
require an appreciation of the subtleties 
involved in the corresponding averaging process~\cite{dh,derrida};
early efforts in this direction~\cite{glaus} have
since been extended and put into a wider perspective~\cite{crisanti,dQ92,sldq}.
We consider a two-dimensional, square-lattice, random-bond Ising model
with a binary distribution of ferromagnetic interaction strengths,
each occurring with equal probability.
For this specific model, the transition temperature is exactly known 
from duality\cite{fisch,kinzel}, so one can be sure that numerical errors
due to imprecise knowledge of the critical point (such as may happen 
{\it e.g.} for site-diluted cases) are absent. 
The only sources of such errors will then be the finite strip width and 
those arising from the averaging process.
However, the former can be controlled by finite-size scaling
theory\cite{fs2,fs1}, while the effects of the latter are reduced by 
studying large enough samples (though this is a subtle point when 
correlation functions are concerned, as seen below).
A previous strip calculation for this model~\cite{sldq} concentrated
on testing for random systems the well-known relation\cite{cardy} between
the critical exponent $\eta$ and the correlation length on a strip at the
critical temperature of the two-dimensional system, as well as on 
extracting the conformal anomaly $c$ (proportional to
the leading finite-width correction to the bulk free energy~\cite{bcn,aff}).
The conclusion was that, within error bars, $\eta =1/4$ and $c = 1/2$
(the pure Ising values) for wide ranges of disorder.
Since $\eta=2-\gamma/\nu$, those results could not be used to test any
disorder dependence of $\gamma$ and $\nu$ separately. 
Here, instead, we resort to numerical derivatives to obtain $\nu$.

We have used long strips of a square lattice, of width $4 \leq L \leq 12$
sites with periodic boundary conditions. 
In order to provide samples that are sufficiently representative of disorder, 
we iterated the transfer matrix\cite{fs2} typically along $10^7$
lattice spacings, meaning much longer strips than those
used in Ref.\onlinecite{sldq}.

At each step, the respective vertical and horizontal bonds between first-neighbour spins $i$ and $j$ were drawn from a probability distribution 
\begin{equation}
 P(J_{ij})= {1 \over 2} ( \delta (J_{ij} -J_0) +  \delta (J_{ij} -rJ_0) ) 
,\ 0 \leq r \leq 1,
\label{eq:1}
\end{equation}
\noindent which ensures\cite{fisch,kinzel} that the critical temperature $\beta_c = 1/k_B T_c$ \ of the corresponding
two-dimensional system is given by
\begin{equation}
\sinh (2\beta_{c} J_{0})\sinh (2\beta_{c}r J_{0}) = 1 \ \ .
\label{eq:2}
\end{equation}
We have used three values of $r$ in calculations: $r=0.5$, $0.25$ and
$0.1$; the two smallest values have been chosen for the purpose of
comparing with recent Monte-Carlo simulations where $\nu$ and
$\gamma$ are evaluated~\cite{kimunp}. 
A wide range of disorder is thus covered.

The procedure for evaluation of the largest Lyapunov exponent $\Lambda_{L}^{0}$ for a strip of width $L$ and length  $N \gg 1$ is well known\cite{glaus,ranmat}.
The average free energy per site is then 
$f_{L}^{\ ave}(T) = - {1 \over L} \Lambda_{L}^{0}$, in units of $k_{B}T$.

From finite-size scaling, the initial susceptibility of a strip at the
critical temperature of the corresponding infinite system, $\chi_L (T_c)$
must vary as~\cite{fs1} :
\begin{equation}
\chi_L (T_c) ={ \partial^{2}  f_{L}^{\ ave}(T_c) \over \partial
h^2}\Biggr|_{h=0}
= L^{\gamma/\nu}\ Q(0)\ \ ,
\label{eq:5}
\end{equation}
\noindent where $h$ is a uniform external field and $Q(0)$ is a 
constant\cite{Fisher71}. As $f_{L}^{\ ave}(T)$ is
expected to have a normal distribution~\cite{derrida,ranmat}, so will
$\chi_L$. Thus the fluctuations are  Gaussian, and relative errors 
must die down with sample size (strip length) $N$ as $1/\sqrt{N}$. 
The intervals (of external field values, in this case) used in  
obtaining finite differences for the calculation of numerical derivatives
must be strictly controlled, so as not to be
an important additional source of errors. We have managed to minimise
these latter effects by using $\delta h$ typically of order $10^{-4}$
in units of $J$ when estimating  $f_{L}^{\ ave}(T_c; h=0,\, \pm \delta h)$
for the derivative in Eq.\ (\ref{eq:5}).

A succession of estimates, $\left(\gamma/\nu\right)_L$, for the 
ratio $\gamma/\nu$ is then obtained from Eq.\ (\ref{eq:5}) as follows:
\begin{equation}
\left({\gamma\over\nu}\right)_L=
{\ln \left[\chi_L(T_c)/\chi_{L-1}(T_c)\right]
\over
\ln \left[L/(L-1)\right]}\\
\label{eq:5p}
\end{equation}
Least-squares fits for plots of $\left(\gamma/\nu\right)_L$ against
$1/L^2$ (see, e.g., Ref.~\onlinecite{sldq} for a discussion
of suitable powers of $1/L$ for extrapolation) provide the 
following results: $\gamma/\nu=1.748\pm 0.012,\ 1.749\pm 0.008,$ and
$1.746\pm 0.013$, respectively for $r=0.50,\ 0.25,$ and 0.10; 
the latter two estimates agree with 
$1.74\pm 0.03,\ 1.73\pm 0.05$, obtained in Ref.\ \cite{kimunp}.

The overall picture is thus consistent with $\gamma/\nu = 7/4$ for all
degrees of disorder. 
Taken together with the results of Ref.~\cite{sldq}, and using the 
scaling relation $\gamma/\nu = 2 - \eta$, this confirms
the view that: 
(1) the conformal invariance relation~\cite{cardy} 
$\eta =L/\pi \xi_{L}(T_c)$
still holds for disordered systems, provided that an averaged correlation
length is used; 
and that (2) the appropriate correlation length to be used is that coming 
from the slope of semi-log plots of correlation functions against
distance~\cite{sldq}. 

We now present results for the exponent $\nu$. The first difference to the
free energy calculation described above is that the correlation functions are
expected to have a {\it log}-normal distribution~\cite{dh,derrida} rather 
than a normal one. Thus self-averaging is not present, and fluctuations for
a given sample do $not$ die down with increasing sample size. However, we have
seen that overall averages ({\it i.e.} central estimates) from different
samples do get closer to each other as the various samples' sizes increase. 
Accordingly, in what follows the error bars quoted arise from fluctuations
among four central estimates, each obtained from a different impurity
distribution. Similar procedures seem to have been followed in
Monte-Carlo calculations of correlation functions in finite ($ L \times L$)
systems~\cite{talapov}.

The direct calculation of correlation functions, 
$\langle\sigma_{0} \sigma_{R}\rangle$, 
follows the lines of Section 1.4 of Ref. \onlinecite{fs2}, with standard 
adaptations for an inhomogeneous system\cite{sldq}. 
For fixed distances up to $R=100$, and for strips with the same length
as those used for averaging the free energy, the correlation functions
are averaged over an ensemble of $10^4$--$10^5$ different estimates 
to yield $\overline{\langle\sigma_{0} \sigma_{R}\rangle}$. 

The average correlation length, $\xi^{ave}$, is in turn defined by
\begin{equation}
\overline{\langle\sigma_{0} \sigma_{R}\rangle}
\sim
\exp\left(-R/\xi^{ave}\right),\\
\label{xi}
\end{equation}
and is calculated from least-squares fits of straight lines to semi-log 
plots of the average correlation function as a function of distance,
in the range $10\leq R\leq 100$. 

We can then apply the usual finite-size scaling (FSS) 
arguments\cite{Fisher71,fs1} to obtain estimates $\nu_L$ of the exponent $\nu$. 
Assuming a simple power-law divergence -- i.e., ignoring, for the time being,
less-divergent terms such as power-law or logarithmic corrections -- 
of the correlation length in the form $\xi\sim t^{-\nu}$, with $t$ being
some reduced distance to the critical point, its FSS {\it ansatz} becomes
\begin{equation}
\xi_L^{ave}=L\ {\cal F}(z),
\label{simple}
\end{equation}
where $z=tL^{1/\nu}$ and ${\cal F}$ is a scaling function. 
Since $\nu$ does not appear explicitly in the expression for
$\xi_L^{ave}(T_c)/L$,
one resorts to the
temperature derivative of the correlation length,
which can also be cast in a similar
scaling form,
\begin{equation}
\mu_L\equiv{d\xi^{ave}_L\over dt}=
L^{1+{1\over\nu}}\ {\cal G}(z),
\label{xip}
\end{equation}
with ${\cal G}\equiv d{\cal F}/dz$. 
$\mu_L$ at $T_c$ (see Eq.\ \ref{eq:2}) is calculated numerically from 
values of $\xi^{ave}_L$ evaluated at $T_c\pm\delta T$, with $\delta T/T_c=10^{-3}$.
For systems of sizes $L$ 
and $L-1$, one obtains the estimates
\begin{equation}
{1 \over \nu_L} = 
{ \ln \left(\mu_L/\mu_{L-1}\right)_{T=T_c}
\over \ln (L/L-1)} - 1\ .
\label{eq:7}
\end{equation}

Note that this is slightly different from the usual fixed-point 
calculation~\cite{Nightingale76,fs2}, and is more convenient in 
the present case where the exact critical temperature is known. 
Our data for each pair of ($L, L-1$) strips are shown in Table I, together
with results of extrapolations against $1/L^2$
for each separate sequence corresponding to different values of $r$.
Taken at face value, the data show a systematic trend towards values of $\nu$
slightly larger than the pure-system value of 1, though the variation is 
smaller than that shown in Ref.~\onlinecite{kimunp}.

Before accepting this trend as an indication of the weak-universality 
scenario, we must test for corrections caused by less-divergent terms
as being responsible for the observed disorder dependence of $\nu$. 
We first recall that logarithmic corrections have already been proposed 
for the bulk correlation length in the form\cite{dotsenko}
\begin{equation}
\xi\sim t^{-\nu}\left[1+C\ln \left(1/t\right)\right]^{\tilde\nu},\\
\label{bulk}
\end{equation}
with $\nu=1$ and $\tilde\nu=1/2$, and $C$ is a (disorder-dependent) constant. 
For the same reasons as above, estimates for $\nu$ can only be
consistently tested
through the temperature derivative of $\xi$:
\begin{equation}
\mu\equiv
{d\xi\over dt}\sim
t^{-(1+\nu)}\left[1+C\ln \left(1/t\right)\right]^{\tilde\nu},\\
\label{bulkp}
\end{equation}
plus less-divergent terms.

For finite systems, logarithmic corrections are expected to 
show 
on scales larger than a disorder-dependent characteristic length 
$L_C \sim \exp(1/C)$~\cite{dotsenko}.
A finite-size scaling {\em ansatz} for the behaviour at $T_c$ 
can be obtained by a suitable generalisation of the standard
procedures for pure power-law singularities (see, e.g., Ref.\onlinecite{fs1}). 
Assuming $\nu=1$ and $\tilde\nu=1/2$ one then has, to dominant order: 
\begin{equation}
{\mu_L\over L^2}\sim \left(1 - b \ln L\right)^{1/2}\ ,\\
\label{strip}
\end{equation}
where $b \sim 1/\ln L_C$. 
Figure \ref{figrb} shows the results for
$\left(\mu_L/L^2\right)^2$ as a function of $\ln L$, for different 
values of $r$. 
In each case, log-corrected behaviour sets in for suitably large $L$,
exactly in the manner predicted by theory: the data stabilize 
onto a straight line only for $L\gtrsim L_C$, which decreases with 
increasing disorder\cite{dotsenko}. 

This crossover effect is similar to that found by Wang 
{\it et al.}\cite{wang}
in their fitting of specific heat data to the double logarithmic
divergence predicted by theory\cite{dotsenko}.
However, the increasing broadening of the specific heat maximum 
with disorder in finite-sized systems has been interpreted as 
evidence against double-logarithmic corrections\cite{kim,comments}. 
Here, instead, we deal with a case of single-log corrections 
to a divergence much stronger than that of the specific heat.
It is thus easier to separate between corrections and the 
dominant power-law behaviour,
as made evident by the consistent fits displayed in Fig.\ \ref{figrb}.
Though $\xi_L(T)$ and the susceptibility $\chi_L(T)$ 
were calculated through Monte Carlo
simulations in Refs.\ \cite{kim,kimunp}, no attempt seems to have been
made to fit the corresponding data to a form similar to Eq.\ (\ref{strip}).
It must be recalled that by examining the behaviour at the exact $T_c$, 
only finite-size effects play any role in tuning the crossover;
by contrast, when $T_c$ is not known exactly, one cannot be sure
whether the (thermal) crossover towards critical behaviour has already 
occurred. As a final check of our data, we have also tried less-divergent 
power-law corrections, but the fittings were always much poorer than those 
assuming logarithmic corrections. In view of these facts, the estimates 
provided by Eq.\ (\ref{eq:7}) should then be regarded as {\em effective} 
exponents, since strong universality still holds.
 
We have also applied the ideas behind Eq.\ (\ref{strip}) to the 
two-dimensional site-diluted Ising model, using the calculational scheme
proposed in Ref.~\onlinecite{dQ92}. 
Results for the pure and diluted cases (for concentrations of magnetic sites
in the range $p=0.65$ -- 0.95) and $L= 3 - 7$ are depicted in Fig.\ \ref{dil}.
It can be seen that the  
qualitative trend clearly changes towards a $\ln L$--dependence similar to 
that found in the random-bond case
as soon as dilution is introduced.
The small curvature in the plots must be at least partly attributed to
imprecise knowledge of the exact critical line (and to the approximate
nature of that calculational scheme itself, which is asymptotically exact only
as $T \to 0$\cite{dQ92}). Again, pure-system exponents with logarithmic corrections 
(as opposed to dilution-dependent ones\cite{kim,kuhn}) seem to 
describe the  behaviour of site-diluted Ising magnets in two dimensions.   

In conclusion, our data independently confirm that the conformal invariance
result $\xi^{ave}=L/\pi\eta$ is still valid for the two-dimensional random--bond
Ising model, with $\eta=1/4$ as in the pure case.
The apparent dependence of $\nu$ with disorder was found to be due 
to logarithmic corrections, which become more important the farther one moves
away from the pure (i.e, $r=1$) system. 
The weak-universality scenario, though quite appealing for the possibility
of demanding new underlying concepts to be explained, does not seem
to hold in the two-dimensional random-bond Ising model. A similar
picture most likely holds for the site-diluted model as well.
The results presented here do not necessarily imply, however, that conformal
invariance
or strong universality should be valid for {\em any} type of disorder.
In the problems treated here correlations
can still freely propagate, unlike cases
where {\it e.g.} frustration is allowed; 
we are currently investigating these issues for spin-glass--like systems.

We thank Laborat\'orio Nacional de Computa\c c\~ao Cien\-t\'\i\-fica 
(LNCC) for use of their computational facilities, and
Brazilian agencies  CNPq
and FINEP,
for financial support.
SLAdQ thanks the Department of Theoretical Physics
at Oxford, where part of this work was done, for the hospitality, and
the cooperation agreement between Academia Brasileira de Ci\^encias and
the Royal Society for funding his visit. Special thanks are due to R. B.
Stinchcombe for invaluable discussions, and to D. Stauffer for useful
suggestions.

\begin{table}
\caption{
Critical exponent $\nu$ from Eq.\ (\protect{\ref{eq:7}}).}
\vskip 0.7cm 
 \halign to \hsize{
\hfil#\quad\hfil&\quad\hfil#\quad\hfil&\hfil#\quad\hfil&\hfil#\quad\hfil\cr
    $L$ & $r=0.50$ &  0.25 &  0.10 \cr \noalign{\smallskip}
  5  & $0.928\pm 0.004$ & $0.993\pm 0.015$& $1.13\pm 0.11$\cr
  6  & $0.962\pm 0.010$ & $1.029\pm 0.024$& $1.15\pm 0.06$\cr
  7  & $0.981\pm 0.020$ & $1.040\pm 0.036$& $1.14\pm 0.06$\cr
  8  & $0.997\pm 0.025$ & $1.053\pm 0.040$& $1.15\pm 0.06$\cr
  9  & $1.000\pm 0.030$ & $1.052\pm 0.026$& $1.13\pm 0.06$\cr
 10  & $1.009\pm 0.033$ & $1.063\pm 0.016$& $1.15\pm 0.07$\cr
 11  & $1.012\pm 0.032$ & $1.062\pm 0.017$& $1.15\pm 0.08$\cr
 12  & $1.016\pm 0.039$ & $1.064\pm 0.032$& $1.14\pm 0.11$\cr
Extrap.& $1.037 \pm 0.016$ & $1.083 \pm 0.014$ & $1.14 \pm 0.06$ \cr
Ref. \protect{\cite{kimunp}} &  & $1.09 \pm 0.01$ & $1.23 \pm 0.02$ \cr
}
\end{table}

\begin{figure}
\caption{Finite-size scaling plots of logarithmic corrections 
[Eq.\ (\protect{\ref{strip}})].
Straight lines are least-squares fits of data respectively for
$L= 9 - 12$ ($r=0.5$); $7 - 12$ ($r=0.25$)  and  $4 - 12$ ($r=0.1$).
For data in this figure, $\mu_L=d\xi_L/dK$, with $K\equiv J/T$.}
\label{figrb}
\end{figure}

\begin{figure}
\caption{Finite-size scaling plots of logarithmic corrections 
[Eq.\ (\protect{\ref{strip}})]
for site-diluted Ising model. Top to bottom: concentration
of magnetic sites = 1.0, 0.95, 0.90, 0.80, 0.70. 0.65.  
See Ref. \protect{\cite{dQ92}} for details on the calculation of $\xi_L$.  
}
\label{dil}
\end{figure}

\end{document}